\documentclass[a4paper,11pt]{article}

\usepackage{pos}

\usepackage{braket}

\title{Non-Hermiticity:~a new paradigm for model building in particle physics}

\author*{Peter Millington}

\affiliation
{School of Physics and Astronomy, University of Nottingham,\\
University Park, Nottingham NG7 2RD, United Kingdom}

\emailAdd{p.millington@nottingham.ac.uk}

\abstract{Non-Hermitian quantum theories have been applied in many other areas of physics. In this note, I will briefly review recent developments in the formulation of non-Hermitian quantum field theories, highlighting features that are unique compared to Hermitian theories. I will touch upon their crucial discrete symmetries and how continuous symmetry properties are borne out, including Noether's theorem, the Goldstone theorem and the Englert-Brout-Higgs mechanism. As examples, I will describe a non-Hermitian supersymmetric theory, and draw attention to non-Hermitian deformations of (scalar) QED, the Higgs-Yukawa theory and flavour oscillations, with potential implications for non-Hermitian model building in the neutrino sector. Together, these results pave the way for a systematic programme for building non-Hermitian extensions of the Standard Model of particle physics.}

\FullConference{
  *** The European Physical Society Conference on High Energy Physics (EPS-HEP2021), ***\\
  *** 26-30 July 2021 ***\\
  *** Online conference, jointly organized by Universität Hamburg and the research center DESY ***
}

\begin{document}
\maketitle


\section{Introduction}

To go beyond the Standard Model of particle physics, we can do two things:\ (i) add new degrees of freedom, e.g., extra Higgs doublets, right-handed neutrinos, SUSY partners, or gauge singlets, perhaps coupling to hidden sectors; or (ii) relax assumptions, e.g., the number of spatial dimensions, Lorentz invariance, locality, or CPT invariance.
The focus of this note is the possibility to relax Hermiticity of the Lagrangian/Hamiltonian.  This option stems from the following observation:\ Hermitian operators have real eigenvalues, but not all operators with real eigenvalues are Hermitian.  Hermiticity of the Hamiltonian is a sufficient not a necessary condition to have a real spectrum~\cite{Bender:1998ke}. Instead, the Hamiltonian must have an antilinear symmetry, under the action of which energy eigenstates form a common eigenbasis~\cite{Bender:1998ke, Mostafazadeh:2001jk}. The argument is as follows (based here on Ref.~\cite{Mannheim:2015hto}):

Given an eigenstate $\ket{\psi(t)}$ of the Hamiltonian $\hat{H}$ with energy $E\in\mathbb{C}$, we have $\hat{H}\ket{\psi(t)}=E\ket{\psi(t)}$. Acting with an antilinear operator $\hat{A}$, such that $\forall\lambda\in\mathbb{C}: \hat{A}\lambda\hat{A}^{-1}=\lambda^*$, and taking $t\to -t$, we obtain $\hat{A}\hat{H}\ket{\psi(-t)}=E^*\ket{\hat{A}\psi(-t)}$. If $\hat{H}$ is invariant under the action of $\hat{A}$, i.e., $\hat{A}\hat{H}\hat{A}^{-1}=\hat{H}$, then $\hat{H}\ket{\hat{A}\psi(-t)}=E^*\ket{\hat{A}\psi(-t)}$. Thus, if $\ket{\hat{A}\psi(-t)}\not{\propto}\ket{\psi(t)}$ then the energies come in complex-conjugate pairs, and if $\ket{\hat{A}\psi(-t)}\propto\ket{\psi(t)}$ then $\hat{H}\ket{\psi(t)}=E^*\ket{\psi(t)}$ and $E=E^*\in\mathbb{R}$.


\section{Scalar prototype}

Let's consider a simple linear, scalar and non-Hermitian field theory with real Lagrangian parameters.  It is composed of two complex scalar fields $\phi_j$ ($j=1,2$) with a non-Hermitian mass mixing. The $c$-number Lagrangian can be written in the following form~\cite{Alexandre:2020gah}:
\begin{equation}
    \label{eq:scalar_Lag_tildes}
    \mathcal{L}_{\phi}=\partial_{\alpha}\tilde{\phi}_j^{*}\partial^{\alpha}\phi_j-M_{jk}^2\tilde{\phi}_j^{*}\phi_k,\qquad 
    M^2=\begin{pmatrix} m_1^2 & \mu^2\\ -\mu^2 & m_2^2\end{pmatrix},\qquad m_1^2,m_2^2,\mu^2,m_1^2-m_2^2>0\,,
\end{equation}
and $\alpha$ is a Lorentz index. The tildes indicate that the $\tilde{\phi}_j^*$ are not simply the complex conjugates of $\phi_j$. The $\phi_j$ and $\tilde{\phi}_j$ equations of motion  differ by $\mu^2\to-\mu^2$, and this ensures that the Euler-Lagrange equations obtained from the Lagrangian~\eqref{eq:scalar_Lag_tildes} are mutually consistent~\cite{Alexandre:2020gah}.

The matrix $M^2$ is real skew symmetric, and therefore not Hermitian. However, its eigenvalues
\begin{equation}
    \label{eq:etadef}
    m_{\pm}^2=\big(m_1^2+m_2^2\big)/2\pm\sqrt{1-\eta^2}\left|m_1^2-m_2^2\right|/2\,,\qquad     \eta=2\mu^2/\big(m_1^2-m_2^2\big)\,,
\end{equation}
are real when the argument of the square root is positive semi-definite. The parameter $\eta$ determines the phase of the antilinear symmetry:\ for $\eta^2=0$, we have a Hermitian theory; for $0\leq \eta^2<1$, the antilinear symmetry is unbroken, and the eigenspectrum is real; and for $\eta^2>1$, the antilinear symmetry is broken, and the eigenspectrum is complex. When $\eta^2=1$, we are at an \emph{exceptional point}, where the eigenvalues merge, i.e., $m_+^2=m_-^2$, $M^2$ becomes defective, and we lose an eigenvector.

If we take one of the fields, $\phi_2$ say, to transform as a pseudoscalar under parity $\mathcal{P}$, the antilinear symmetry guaranteeing the reality of the eigenvalues is symmetry under the combined action of parity $\mathcal{P}$ and time-reversal $\mathcal{T}$, which act on the $c$-number fields as follows:
\begin{equation}
    \mathcal{P}:\ \phi_j(x)\to \phi'_j(\mathcal{P}x)=P_{jk}\tilde{\phi}_k(x)\,,\qquad
    \mathcal{T}:\ \phi_j(x)\to \phi'_j(\mathcal{T}x)=\phi^*_j(x)\,,
\end{equation}
wherein $P=P^{-1}={\rm diag}(1,-1)$ is the parity matrix. Notice that parity relates the untilded and tilded fields, since $P\cdot M^2\cdot P=\left(M^2\right)^{\mathsf{T}}$, where $\mathsf{T}$ denotes the matrix transpose, effecting $\mu^2\to-\mu^2$.

An alternative description of this model was first introduced in Ref.~\cite{Alexandre:2017foi}, with the Lagrangian
\begin{equation}
    \label{eq:scalar_Lag}
    \mathcal{L}_{\phi}'=\partial_{\alpha}\phi_j^{*}\partial^{\alpha}\phi_j-m_j^2\phi_j^{*}\phi_j-\mu^2\left(\phi_1^{*}\phi_2-\phi_2^{*}\phi_1\right).
\end{equation}
Note the absence of tildes on the conjugated fields. This description has the feature that the equations of motion obtained by varying the action with respect to $\phi_j$ and $\phi^*_j$ are not mutually consistent~\cite{Alexandre:2017foi}. However, by choosing the equations of motion by varying with respect to either the $\phi_j$ or the $\phi_j^*$~\cite{Alexandre:2017foi}, the resulting dynamics is identical to that obtained from varying Eq.~\eqref{eq:scalar_Lag_tildes}. The exception, however, is in the way that Noether's theorem is borne out, with conserved currents arising for transformations that effect particular variations of the non-Hermitian part of the Lagrangian~\eqref{eq:scalar_Lag}~\cite{Alexandre:2017foi}. 

The eigenvectors of the mass matrix are $\mathbf{e}_+=N\big(\eta, -1+\sqrt{1-\eta^2}\big)$ and $\mathbf{e}_-=N\big(-1+\sqrt{1-\eta^2}, \eta\big)$, where $N$ is some normalization~\cite{Alexandre:2017foi}. However, since the mass matrix is not normal for $\mu^2\neq 0$ (i.e., \smash{$[M^2,(M^2)^{\dag}]\neq 0$}), except when $m_1^2=m_2^2$, it cannot be diagonalized by a unitary transformation and its eigenvectors are not orthogonal with respect to the usual Hermitian inner product.  The mass matrix can, however, be diagonalized by a similarity transformation~\cite{Alexandre:2017foi} (see also Ref.~\cite{Alexandre:2020gah}):\footnote{The eigenvalues would be interchanged were we to take $m_2^2-m_1^2>0$, while keeping $\eta$ as defined in Eq.~\eqref{eq:etadef}, such that $R\cdot M^2\cdot R^{-1}={\rm diag}(m_-^2,m_+^2)$ for $R$ as given in Eq.~\eqref{eq:similarity}.}
\begin{equation}
    \label{eq:similarity}
    R\cdot M^2\cdot R^{-1}=\begin{pmatrix} m_+^2 &  0 \\ 0 & m_-^2\end{pmatrix},\qquad 
    R=N\begin{pmatrix} \eta & 1-\sqrt{1-\eta^2} \\ 1-\sqrt{1-\eta^2} & \eta\end{pmatrix}.
\end{equation}
The existence of this similarity transformation allows us to define an inner product under which the eigenvectors are orthonormal and their norms are positive (see Refs.~\cite{Mostafazadeh:2001jk, Bender:2002vv}): $\mathbf{e}^*_{\pm}\cdot R^2\cdot\mathbf{e}_{\mp}=0$ and $\mathbf{e}^*_{\pm}\cdot R^2\cdot\mathbf{e}_{\pm}=1$, with \smash{$N=\big(2\eta^2-2+2\sqrt{1-\eta^2}\big)^{-1/2}$}. Notice that $|N|\to\infty$ at the exceptional points ($\eta^2\to 1$). Since $R\cdot R=R\cdot P\cdot R^{-1}\cdot P$, we can isolate a matrix $C'=R\cdot P\cdot R^{-1}$ that is associated with an additional symmetry of the Lagrangian (introduced in Ref.~\cite{Bender:2002vv} as the ``$\mathcal{C}$'' symmetry), which acts at the level of the squared mass matrix as $C^{\prime\mathsf{T}}\cdot M^2\cdot C^{\prime\mathsf{T}}=M^2$ and plays a role in ensuring unitary evolution~\cite{Bender:2002vv}. Even so, to describe non-Hermitian flavour oscillations, we must consider physical scattering processes to avoid apparent problems with perturbative unitarity~\cite{Alexandre:2020gah} (cf.~Ref.~\cite{Ohlsson:2019noy}).


\section{Spontaneous symmetry breaking (SSB)}

Within the regime of unbroken $\mathcal{PT}$ symmetry, the existence of a conserved current is sufficient to ensure that the Goldstone theorem continues to hold in the case of a spontaneously broken global symmetry~\cite{Alexandre:2018uol}. For instance, for the non-Hermitian model
\begin{equation}
    \mathcal{L}_{\rm SSB}=\partial_{\alpha}\phi_j^*\partial^{\alpha}\phi_j+m_1^2|\phi_1|^2-m_2^2|\phi_2|^2-\mu^2\left(\phi_1^*\phi_2-\phi_2^*\phi_1\right)-\tfrac{g}{4}|\phi_1|^4\,,\qquad m_1^2,m_2^2,\mu^2,g>0\,,
\end{equation}
we can expand around the minima determined via
\begin{equation}
    \left.\begin{matrix}\left.\frac{\partial \mathcal{L}}{\partial \phi_1^*}\right|_{\phi_j=v_j}=m_1^2v_1-\mu^2v_2-\frac{g}{2}|v_1|^2v_1=0\\ \left.\frac{\partial \mathcal{L}}{\partial \phi_2^*}\right|_{\phi_j=v_j}=-m_2^2v_2+\mu^2v_1=0\end{matrix}\right\}\Rightarrow\begin{pmatrix} v_1 \\ v_2\end{pmatrix}=\sqrt{\frac{2}{g}\frac{m_1^2m_2^2-\mu^4}{m_2^2}}\,e^{i\vartheta}\,,\qquad \vartheta \in\mathbb{C}\,,
\end{equation}
and take $\phi_j\to v_j+\delta\phi_j$ to show that there exists a Goldstone mode $G_0\propto {\rm Im}\,\delta \phi_1-(\mu^2/m_2^2)\,{\rm Im}\,\delta \phi_2$~\cite{Alexandre:2018uol}.

We can gauge the U(1) symmetry of the above model, and the Englert-Brout-Higgs mechanism is still borne out. For instance, taking ($F_{\alpha\beta}=\partial_{\alpha}A_{\beta}-\partial_{\beta}A_{\alpha}$ and $D_{\alpha}=\partial_{\alpha}+iqA_{\alpha}$, with $q\in \mathbb{R}$)
\begin{equation}
    \mathcal{L}_{{\rm SSB},A}=-\tfrac{1}{4}F_{\alpha\beta}F^{\alpha\beta}+D^*_{\alpha}\phi_j^*D^{\alpha}\phi_j+m_1^2|\phi_1|^2-m_2^2|\phi_2|^2-\mu^2\left(\phi_1^*\phi_2-\phi_2^*\phi_1\right)-\tfrac{g}{4}|\phi_1|^4-\tfrac{1}{2\xi}(\partial_{\alpha}A^{\alpha})^2\,,
\end{equation}
the squared gauge boson mass after SSB is $m_A^2=2q^2(|v_1|^2+|v_2|^2)$~\cite{Alexandre:2018xyy} (see Ref.~\cite{Alexandre:2019jdb} for the non-Abelian case). Interestingly, this disagrees with alternative treatments based on similarity transformations to Hermitian theories~\cite{Mannheim:2018dur, Fring:2019hue}. The latter effects $|v_2|^2\to -|v_2|^2$ in $m_A^2$, and the gauge boson mass then vanishes at the exceptional points ($\mu^2=\pm m_2^2$), despite the broken symmetry.


\section{Non-Hermitian Dirac theory}

We can make a non-Hermitian deformation of the theory of a single Dirac fermion $\psi$ by introducing a parity odd, anti-Hermitian mass term~\cite{Bender:2005hf}:
\begin{equation}
    \label{eq:Dirac}
    \mathcal{L}_{\rm Dir}=\bar{\psi}(i\gamma^{\alpha}\partial_{\alpha}-m-\mu\gamma^5)\psi\,,\qquad m,\mu\in\mathbb{R}\,,
\end{equation}
where $\gamma^5=\left(\gamma^5\right)^{\dag}$ is the fifth gamma matrix. The energies are given by the roots of $E^2=\mathbf{p}^2+m^2-\mu^2$, and there exists a conserved current, which is given by~\cite{Alexandre:2015oha} (see also Ref.~\cite{Alexandre:2017foi})
\begin{equation}
    j^{\alpha}=\bar{\psi}\gamma^{\alpha}\big(1+\mu\gamma^5/m\big)\psi=\bar{\psi}\gamma^{\alpha}\left[\left(1-\mu/m\right)P_L+\left(1+\mu/m\right)P_R\right]\psi\,,
\end{equation}
where $P_{R(L)}=\big(\mathbb{I}+(-)\gamma^5\big)/2$ are the right- ($R$) and left-chiral ($L$) projection operators.

At the exceptional points $\mu=\pm m$, the theory behaves on-shell as that of a single massless left- or right-chiral fermion~\cite{Alexandre:2015kra} (see similar behaviour on the lattice~\cite{Chernodub:2017lmx}). In fact, adding an axial vector gauge coupling, the gauge symmetry is recovered at the exceptional points~\cite{Alexandre:2015kra}, despite the presence of mass terms in the Lagrangian. The persistence of these mass terms at the exceptional points also means that massless fermions may undergo flavour oscillations~\cite{Jones-Smith:2009qeu}, with potential implications for the neutrino sector. Additionally, the structure of Hermitian and anti-Hermitian mass terms in Eq.~\eqref{eq:Dirac} can be obtained from a non-Hermitian Higgs-Yukawa theory~\cite{Alexandre:2015kra} (see also Ref.~\cite{Alexandre:2020bet}).


\section{Supersymmetry (SUSY) embedding}

The non-Hermitian models described above are composed of four degrees of freedom:  we have two flavours of complex scalar field or two chiral components of a Dirac fermion. The former of these can be partnered with two flavours of Majorana fermion, and both the fermion and scalar components can be packaged within two $\mathcal{N}=1$ scalar chiral superfields $\Phi_j$ ($j=1,2$). We can then write a non-Hermitian supersymmetric Lagrangian~\cite{Alexandre:2020wki}
\begin{subequations}
\begin{gather}
    \mathcal{L}_{\rm SUSY}=\int{\rm d}^2\theta^{\dag}\,{\rm d}^2\theta\,\left(|\Phi_1|^2+|\Phi_2|^2\right)+\int{\rm d}^2\theta\, W_++\int{\rm d}^2\theta^{\dag}\, W_-^{\dag}\,,\\
    W_{\pm}=\tfrac{1}{2}m_{11}\Phi_1^2\mp m_{12}\Phi_1\Phi_2+\tfrac{1}{2}m_{22}\Phi_2^2\,,
\end{gather}
\end{subequations}
where $\theta$ and $\theta^{\dag}$ are Grassmann variables. On-shell, this yields the scalar model~\eqref{eq:scalar_Lag}, with the squared mass parameters given by $m_j^2=m_{jj}^2-m_{12}^2$ and $\mu^2=m_{12}(m_{22}-m_{11})$, along with a system of Majorana fermions described by the Lagrangian
\begin{equation}
    \mathcal{L}_{\rm Maj}=\tfrac{1}{2}\bar{\psi}_j i \gamma^{\alpha}\partial_{\alpha}\psi_j-\tfrac{1}{2}m_{jj}\bar{\psi}_j\psi_j-\tfrac{1}{2}m_{12}(\bar{\psi}_1\gamma^5\psi_2+\bar{\psi}_2\gamma^5\psi_1)\,.
\end{equation}
The squared mass eigenvalues are
\begin{equation}
    m_{\pm,s}^2=\big(m_{11}^2+m_{22}^2\big)/2-m_{12}^2\pm\sqrt{\big(m_{11}^2-m_{22}^2\big)^2/4-m_{12}^2\left(m_{11}-s m_{22}\right)^2}\,,\\
\end{equation}
where $s=+1$ for the bosonic fields and $s=-1$ for the fermionic fields, with their spectra agreeing only in the Hermitian limit $m_{12}\to 0$. The non-Hermitian terms therefore provide a source of supersymmetry breaking, despite the theory itself being written entirely in terms of superfields~\cite{Alexandre:2020wki}.


\section{Concluding remarks}

We have highlighted how antilinear symmetries can make certain non-Hermitian quantum theories viable, described a number of prototype non-Hermitian field theories, and considered their continuous symmetry properties.  The key import is that non-Hermitian quantum field theories may provide new avenues for model building in particle physics that exhibit unique phenomenology, e.g., through the existence of exceptional points, which cannot be replicated by Hermitian theories.


\acknowledgments

PM would like to thank Jean Alexandre, Carl M.~Bender, Maxim N.~Chernodub, Edmund J.~Copeland, John Ellis, Esra Sablevice and Dries Seynaeve for enjoyable collaboration in this area. This work was supported by a United Kingdom Research and Innovation (UKRI) Future Leaders Fellowship [Grant No.~MR/V021974/1]; and a Nottingham Research Fellowship from the University of Nottingham.



\begin{thebibliography}{99}

\bibitem{Bender:1998ke}
    C.~M.~Bender and S.~Boettcher,
    \emph{Real spectra in non-Hermitian Hamiltonians having $\mathcal{PT}$ symmetry},
    \href{https://doi.org/10.1103/PhysRevLett.80.5243}{Phys.\ Rev.\ Lett.\ \textbf{80} (1998) 5243}
    \href{https://arxiv.org/abs/physics/9712001}{[physics/9712001]}.

\bibitem{Mostafazadeh:2001jk}
    A.~Mostafazadeh,
    \emph{Pseudo-Hermiticity versus PT symmetry:\ The necessary condition for the reality of the spectrum of a non-Hermitian Hamiltonian}, \href{https://doi.org/10.1063/1.1418246}{J.\ Math.\ Phys.\ \textbf{43} (2002) 205}
    \href{https://arxiv.org/abs/math-ph/0107001}{[math-ph/0107001]};
    \emph{Pseudo-Hermiticity versus PT-symmetry.~II.~A complete characterization of non-Hermitian Hamiltonians with a real spectrum},
    \href{https://doi.org/10.1063/1.1461427}{ibid.\ 2814}
    \href{https://arxiv.org/abs/math-ph/0110016}{[math-ph/0110016]};
    \emph{Pseudo-Hermiticity versus PT-symmetry III:\ Equivalence of pseudo-Hermiticity and the presence of antilinear symmetries},
    \href{https://doi.org/10.1063/1.1489072}{ibid.\ 3944}
    \href{https://arxiv.org/abs/math-ph/0203005}{[math-ph/0203005]}.
    
\bibitem{Mannheim:2015hto}
    P.~D.~Mannheim,
    \emph{Antilinearity rather than Hermiticity as a guiding principle for quantum theory},
    \href{https://doi.org/10.1088/1751-8121/aac035}{J.\ Phys.\ A \textbf{51} (2018) 315302}
    \href{https://arxiv.org/abs/1512.04915}{[1512.04915 [hep-th]]}.

\bibitem{Alexandre:2020gah}
    J.~Alexandre, J.~Ellis and P.~Millington,
    \emph{Discrete spacetime symmetries and particle mixing in non-Hermitian scalar quantum field theories},
    \href{https://doi.org/10.1103/PhysRevD.102.125030}{Phys.\ Rev.\ D \textbf{102} (2020) 125030}
    \href{https://arxiv.org/abs/2006.06656}{[2006.06656 [hep-th]]}.
    
\bibitem{Alexandre:2017foi}
    J.~Alexandre, P.~Millington and D.~Seynaeve,
    \emph{Symmetries and conservation laws in non-Hermitian field theories},
    \href{https://doi.org/10.1103/PhysRevD.96.065027}{Phys.\ Rev.\ D \textbf{96} (2017) 065027}
    \href{https://arxiv.org/abs/1707.01057}{[1707.01057 [hep-th]]}.
    
\bibitem{Bender:2002vv}
    C.~M.~Bender, D.~C.~Brody and H.~F.~Jones,
    \emph{Complex extension of quantum mechanics},
    \href{https://doi.org/10.1103/PhysRevLett.89.270401}{Phys.\ Rev.\ Lett.\ \textbf{89} (2002) 270401} [err.~\href{https://doi.org/10.1103/PhysRevLett.92.119902}{ibid.~\textbf{92} (2004) 119902}]
    \href{https://arxiv.org/abs/quant-ph/0208076}{[quant-ph/0208076]}.
    
\bibitem{Ohlsson:2019noy}
    T.~Ohlsson and S.~Zhou,
    \emph{Transition probabilities in the two-level quantum system with PT-symmetric non-Hermitian Hamiltonians},
    \href{https://doi.org/10.1063/5.0002958}{J.\ Math.\ Phys.\ \textbf{61} (2020) 052104}
    \href{https://arxiv.org/abs/1906.01567}{[1906.01567 [quant-ph]]};
    \emph{Density matrix formalism for $\mathcal{PT}$-symmetric non-Hermitian Hamiltonians with the Lindblad equation},
    \href{https://doi.org/10.1103/PhysRevA.103.022218}{Phys.\ Rev.\ A \textbf{103} (2021) 022218}
    \href{https://arxiv.org/abs/2006.02445}{[2006.02445 [quant-ph]]}

\bibitem{Alexandre:2018uol}
    J.~Alexandre, J.~Ellis, P.~Millington and D.~Seynaeve,
    \emph{Spontaneous symmetry breaking and the Goldstone theorem in non-Hermitian field theories},
    \href{https://doi.org/10.1103/PhysRevD.98.045001}{Phys.\ Rev.\ D \textbf{98} (2018) 045001}
    \href{https://arxiv.org/abs/1805.06380}{[1805.06380 [hep-th]]}.
    
\bibitem{Alexandre:2018xyy}
    J.~Alexandre, J.~Ellis, P.~Millington and D.~Seynaeve,
    \emph{Gauge invariance and the Englert-Brout-Higgs mechanism in non-Hermitian field theories},
    \href{https://doi.org/10.1103/PhysRevD.99.075024}{Phys.\ Rev.\ D \textbf{99} (2019) 075024}
    \href{https://arxiv.org/abs/1808.00944}{[1808.00944 [hep-th]]};

\bibitem{Alexandre:2019jdb}
    J.~Alexandre, J.~Ellis, P.~Millington and D.~Seynaeve,
    \emph{Spontaneously breaking non-Abelian gauge symmetry in non-Hermitian field theories},
    \href{https://doi.org/10.1103/PhysRevD.101.035008}{Phys.\ Rev.\ D \textbf{101} (2020) 035008}
    \href{https://arxiv.org/abs/1910.03985}{[1910.03985 [hep-th]]}.
    
\bibitem{Mannheim:2018dur}
    P.~D.~Mannheim,
    \emph{Goldstone bosons and the Englert-Brout-Higgs mechanism in non-Hermitian theories},
    \href{https://doi.org/10.1103/PhysRevD.99.045006}{Phys.\ Rev.\ D \textbf{99} (2019) 045006}
    \href{https://arxiv.org/abs/1808.00437}{[1808.00437 [hep-th]]}.

\bibitem{Fring:2019hue}
    A.~Fring and T.~Taira,
    \emph{Goldstone bosons in different PT-regimes of non-Hermitian scalar quantum field theories},
    \href{https://doi.org/10.1016/j.nuclphysb.2019.114834}{Nucl.\ Phys.\ B \textbf{950} (2020) 114834}
    \href{https://arxiv.org/abs/1906.05738}{[1906.05738 [hep-th]]};
    \emph{Pseudo-Hermitian approach to Goldstone's theorem in non-Abelian non-Hermitian quantum field theories},
    \href{https://doi.org/10.1103/PhysRevD.101.045014}{Phys.\ Rev.\ D \textbf{101} (2020) 045014}
    \href{https://arxiv.org/abs/1911.01405}{[1911.01405 [hep-th]]};
    \emph{Massive gauge particles versus Goldstone bosons in non-Hermitian non-Abelian gauge theory},
    \href{https://arxiv.org/abs/2004.00723}{2004.00723 [hep-th]}.

\bibitem{Bender:2005hf}
    C.~M.~Bender, H.~F.~Jones and R.~J.~Rivers,
    \emph{Dual $\mathcal{PT}$-symmetric quantum field theories},
    \href{https://doi.org/10.1016/j.physletb.2005.08.087}{Phys.\ Lett.\ B \textbf{625} (2005) 333}
    \href{https://arxiv.org/abs/hep-th/0508105}{[hep-th/0508105]}.
    
\bibitem{Alexandre:2015oha}
    J.~Alexandre and C.~M.~Bender,
    \emph{Foldy--Wouthuysen transformation for non-Hermitian Hamiltonians},
    \href{https://doi.org/10.1088/1751-8113/48/18/185403}{J.\ Phys.\ A \textbf{48} (2015) 185403}
    \href{https://arxiv.org/abs/1501.01232}{[1501.01232 [hep-th]]}.
    
\bibitem{Alexandre:2015kra}
    J.~Alexandre, C.~M.~Bender and P.~Millington,
    \emph{Non-Hermitian extension of gauge theories and implications for neutrino physics},
    \href{https://doi.org/10.1007/JHEP11(2015)111}{JHEP \textbf{11} (2015) 111}
    \href{https://arxiv.org/abs/1509.01203}{[1509.01203 [hep-th]]}.
    
\bibitem{Chernodub:2017lmx}
    M.~N.~Chernodub,
    \emph{The Nielsen--Ninomiya theorem, $\mathcal{PT}$-invariant non-Hermiticity and single 8-shaped Dirac cone},
    \href{https://doi.org/10.1088/1751-8121/aa809a}{J.\ Phys.\ A \textbf{50} (2017) 385001}
    \href{https://arxiv.org/abs/1701.07426}{[1701.07426 [cond-mat.mes-hall]]}.
    
\bibitem{Jones-Smith:2009qeu}
    K.~Jones-Smith and H.~Mathur,
    \emph{Relativistic non-Hermitian quantum mechanics},
    \href{https://doi.org/10.1103/PhysRevD.89.125014}{Phys.\ Rev.\ D \textbf{89} (2014) 125014} \href{https://arxiv.org/abs/0908.4257}{[0908.4257 [hep-th]]}.
    
\bibitem{Alexandre:2020bet}
    J.~Alexandre and N.~E.~Mavromatos,
    \emph{On the consistency of a non-Hermitian Yukawa interaction},
    \href{https://doi.org/10.1016/j.physletb.2020.135562}{Phys.\ Lett.\ B \textbf{807} (2020) 135562}
    \href{https://arxiv.org/abs/2004.03699}{[2004.03699 [hep-ph]]}.

\bibitem{Alexandre:2020wki}
    J.~Alexandre, J.~Ellis and P.~Millington,
    \emph{$\mathcal{PT}$-symmetric non-Hermitian quantum field theories with supersymmetry},
    \href{https://doi.org/10.1103/PhysRevD.101.085015}{Phys.\ Rev.\ D \textbf{101} (2020) 085015}
    \href{https://arxiv.org/abs/2001.11996}{[2001.11996 [hep-th]]}.

\end{thebibliography}
\end{document}